\documentclass[final, 3p, times]{elsarticle}

\usepackage{amssymb}
\usepackage[normalem]{ulem}
\usepackage{xcolor}

\newcounter{bla}

\journal{Computer Physics Communications}

\usepackage{amsmath}

\usepackage{algorithmic}

\usepackage{array}

\usepackage{graphicx}

\begin{document}

\begin{frontmatter}

\title{Towards Lattice Quantum Chromodynamics on FPGA devices}

\author[a]{Grzegorz Korcyl\corref{author}}
\author[b,c]{Piotr Korcyl}

\cortext[author] {Corresponding author.\\\textit{E-mail address:} grzegorz.korcyl@uj.edu.pl}
\address[a]{Department of Information Technologies,  Faculty of Physics, Astronomy and Applied Computer Science, Jagiellonian University, ul. \L ojasiewicza 11, 30-348 Krak\'ow, Poland}
\address[b]{M. Smoluchowski Institute of Physics,  Faculty of Physics, Astronomy and Applied Computer Science, Jagiellonian University, ul. \L ojasiewicza 11, 30-348 Krak\'ow, Poland}
\address[c]{Institut f\"ur Theoretische Physik, Universit\"at Regensburg, 93040 Regensburg, Germany}

\begin{abstract}
In this paper we describe a single-node, double precision Field Programmable Gate Array (FPGA) implementation of the Conjugate Gradient algorithm in the context of Lattice Quantum Chromodynamics. As a benchmark of our proposal we invert numerically the Dirac-Wilson operator on a 4-dimensional grid 
on three Xilinx hardware solutions: Zynq Ultrascale+ evaluation board, the Alveo U250 accelerator and the largest device available on the market, the VU13P device. In our implementation we separate software/hardware parts in such a way that the entire multiplication by the Dirac operator is performed
in hardware, and the rest of the algorithm runs on the host. We find out that the FPGA implementation can offer a performance comparable with that obtained using current CPU or Intel's many core Xeon Phi accelerators. A possible multiple node FPGA-based system is discussed and we argue that power-efficient High Performance Computing (HPC) systems can be implemented using FPGA devices only.
\end{abstract}


\end{frontmatter}

\section{Introduction}

In the last years Field Programmable Gate Array (FPGA) devices started to pave their way into the realm of High Performance Computing (HPC), for examples see \cite{fpga_and_hpc1,fpga_and_hpc2,fpga_and_hpc3}. A
well-known scientific application in this domain are Monte Carlo simulations of Quantum Chromodynamics (QCD), which are performed in the context of
theoretical elementary particles physics. An iterative solver of a large sparse system of linear equations, the Dirac-Wilson operator, lies at the heart of such simulations and is the most compute-intensive kernel. In this work we describe our attempt to port the Conjugate Gradient algorithm \cite{cg} 
which is an example of the most naive of such iterative solvers to a single FPGA device and demonstrate that FPGA devices can compete with modern HPC solutions as far as such academic applications are concerned.

\subsection{QCD Monte Carlo simulations}

From an algorithmic point of view Monte Carlo simulations of Quantum Chromodynamics boil down to a numerical
estimation of highly dimensional integrals. A set of complex numbers representing
the values of basic degrees of freedom: gluon and quark fields, is associated respectively to each edge and point of the space-time lattice. An implementation of a Markov chain allows to construct 
representative candidates of possible configurations of gluon fields according to a desired probability distribution. 
Any function of fundamental quark and gluon variables can be evaluated on each such configuration and the estimate of the final result is obtained
by taking an algebraic average over the estimates from all generated configurations. The larger is the set of available configurations the smaller are the statistical uncertainties and the more precise the final physical result is. 

As long as the probability distribution depends only on gluon fields, the Markov chain update can be performed locally and efficient algorithms 
allow to overcome the critical slowing down reducing the computational complexity to a manageable level. Once the quark field dependence is included 
in the probability distribution, in each update step the determinant of the Dirac operator's matrix has to be estimated and the only applicable algorithm 
in such a situation is the Hybrid Monte Carlo algorithm \cite{DUANE1987216, Urbach:2005ji}. Instead of the 
determinant it is more convenient to estimate the inverse of the Dirac operator using an 
appropriate iterative solver. In many practical cases 90\% of computing time is spent in that solver. 
It is thus crucial to optimize and fully adapt that part of the code to the micro-architecture 
the simulations are running on. 

\subsection{Search for new architectures}

In the physically most interesting cases the Dirac matrix can achieve dimensions of  order $10^9 \times 10^9$ and therefore a distributed, high
performance system is required to handle such a task in a reasonable amount of human time. Various HPC solutions are being used with a particular 
emphasis on fast interconnects as frequent local binary exchanges as well as global operations are required. On typical processors the single-node problem is memory bandwidth bound and hence a fast and large cache memory is desirable. 

Since several years the HPC field witnesses a major shift in the paradigm of the micro-architecture employed, as more and more many-core processors and accelerators are being used. Modern FPGA devices are being considered as candidates for next generation of HPC solutions and some computing centers have already opened FPGA clusters to general public \cite{fpga_cluster}. A dedicated pure FPGA cluster running in production mode for theoretical chemistry application was reported in \cite{aruz}. It is natural to ask whether other traditional HPC applications, such as the Monte Carlo simulations of QCD, can benefit from these new computing resources. In order to answer that question we investigate the performance achieved on a FPGA device by 
the naive solver, the Conjugate Gradient algorithm. Its numerical complexity is directly proportional to the numerical cost of multiplying a vector 
by the Dirac matrix. This matrix-vector multiplication appears ultimately in any other more involved solver, so it is a valid benchmark, while keeping 
the setup as simple as possible.

\subsection{Outline}

The paper is composed as follows. We start by briefly introducing the basic ingredients of lattice QCD, the most important being 
the Dirac matrix which is defined in section \ref{sec. dirac}. For the sake of completeness we present the standard Conjugate Gradient algorithm in 
section \ref{sec. cg} and describe the structure of input and output data in section \ref{sec. input/output}. In section \ref{sec. pld} we provide a brief
overview of the main advantages of the FPGA technology, highlighting concepts which are crucial from the high performance perspective. In section \ref{sec. previous work} we describe previous work in this subject.
In section \ref{sec. design_partition} we pay special attention to the way the part which runs on the host and the part implemented in the logic are separated. We discuss various ways of enforcing the parallelization of the main kernel function in the hardware in \ref{sec. main}. Section \ref{sec. benchmarks}, where we describe our methodology and the details of our runs, is devoted to benchmark results. We focus on the resource utilization (\ref{sec. resources}), problem size limits (\ref{sec. size limits})and time to solution benchmarks (\ref{sec. timings}). We end with conclusions and a discussion of further research directions in section \ref{sec. conclusions}.

\section{Monte Carlo simulations}

\subsection{Lattice Quantum Chromodynamics}
\label{sec. dirac} 

Lattice QCD is defined on a four-dimensional grid of sites, typically with periodic boundary conditions, representing the discretized space-time. 
The basic ingredients are the \emph{gluon (gauge) fields} described by four $3\times3$ complex, unitary matrices $U_{\mu}^{AB}(n)$, $A,B=1,2,3$, $\mu=0,\dots,3$ 
sitting on each forward edge emanating from each site of the lattice, and a \emph{fermion (spinor) field} $\psi_{\alpha}^A(n)$ described by a 12 component 
complex vector living on the lattice sites, with $A=1,2,3$ and $\alpha=0,\dots,3$. The 12 components can be grouped into four sets of three, with the 
gauge matrices rotating each three components independently.

The Dirac operator acting on a spinor vector is defined as \cite{PhysRevD.10.2445} (for a textbook introduction see \cite{gattringer})
\begin{equation}
\psi_{\alpha}^A(m) = D(m,n)^{AB}_{\alpha \beta} \psi_{\beta}^B(n) =  \psi_{\alpha}^A(n) +
 \kappa \sum_{\mu=0}^3 \Big[ U_{\mu}^{AB}(n) P^{-\mu}_{\alpha \beta} \psi_{\beta}^B(n+\hat{\mu}) +
U^{\dagger, AB}(n-\hat{\mu}) P^{+\mu}_{\alpha \beta} \psi_{\beta}^B(n-\hat{\mu}) \Big],
\label{eq. dirac op}
\end{equation}
where \begin{equation}
    \kappa = \frac{1}{2} \frac{1}{m_q + 4}
\end{equation}

and the convention that two indices are always summed over in implied. Moreover
\begin{equation}
P^{\pm \mu} = 1 \pm \gamma_{\mu}.
\end{equation}
As implied by the notation, the $P^{\pm \mu}$ matrices act only on the $\alpha$ index of the spinor field, whereas the $U$ matrices act only on the $A$ index.
The Dirac matrix is sparse as it involves only the nearest-neighbors interactions.

One usually exploits the fact that due to the specific structure of the $P^{\pm \mu}$ matrices, the two lower $\alpha$ components of the spinor are related to the upper two by a simple complex
rescaling. Thus it is sufficient to multiply by the matrix $U$ only the upper components, halving the number of required multiplications by the $U$ matrix. 

\subsection{Conjugate Gradient algorithm}
\label{sec. cg}

The main problem is to solve a set of coupled linear equations described by the abbreviated equation 
\begin{equation}
D^{AB}_{\alpha \beta}(n,m) \psi_B^{\beta}(m) = \eta_A^{\alpha}(n).
\label{eq. dirac eq}
\end{equation}
If $D$ is hermitian and positive-definite a simple iterative conjugate gradient method can be applied. The Dirac operator as defined in Eq. \ref{eq. dirac op} 
satisfies the $\gamma_5$-hermiticity, $\gamma_5 D \gamma_5 = D^{\dagger}$, so one needs to solve for $D D^{\dagger}$ which is hermitian and 
then multiply the solution by $D^{\dagger}$.

The conjugate gradient algorithm which we have implemented reads \cite{cg}
\begin{algorithmic}
\STATE $\psi \gets \psi_0$
\STATE $r \gets \eta - D \psi$
\STATE $p \gets r$
\WHILE {$|r| \geq r_{min}$} 
	\STATE $r_{old} \gets |r|$
        \STATE $\alpha \gets \frac{r_{old}}{|D^{\dagger} p|}$
        \STATE $\psi \gets \psi + \alpha p$
	\STATE $r \gets r - \alpha D D^{\dagger} p$

	\STATE $\beta \gets \frac{|r|}{r_{old}}$
	\STATE $p \gets r + \beta p$
\ENDWHILE
\label{alg. cg}
\end{algorithmic}
All operations on the vectors $r$, $p$ and $\psi$ such as matrix-vector multiplication as well as the computation of the norm $| \cdot |$ involve 
a sum over the entire lattice. The number of iterations before the required residuum is reached depends on the condition number of the Dirac matrix
which in turn depends on the set of all $U$ matrices together with the quark mass $m_q$ and some other parameters.

\subsection{Input and output data}
\label{sec. input/output}

In a typical situation the set of $U$ matrices for the entire lattice resides in the main memory (usually it is read from disc in the initialization phase of the program). A particular physical question can be answered by choosing appropriate 
$\eta$ vector. In the simplest situation it has a single non-zero entry set to one. The CG algorithm above returns then the vector $\psi$ being the solution of the equation Eq. \ref{eq. dirac eq}. Depending
on the circumstances the next computation can involve the same $U$ matrices but a different $\eta$ vector, or, more frequently, a completely new set of $U$ and $\eta$ variables.

In a multi-process implementation the lattice is divided into smaller hybercubes and each hypercube is associated to a process. Multiplication by the Dirac operator Eq. \ref{eq. dirac op}
is performed in each process independently and in parallel, while the global steps of the CG algorithm, such as the evaluation of the stopping criterion are executed by the master process. Hence,
apart from send/receive operations between neighboring processes, frequent global operations such as gather and broadcast are required. Because of the nearest-neighbor structure of the Dirac operator a copy of the boundary values has to be maintained on adjacent processes. For physical reasons periodic or antiperiodic boundary conditions are imposed on the $U$ matrices.  

\subsection{Single stencil calculation}

The most elementary computational block is the evaluation of the single stencil, i.e. evaluation of the right hand side of Eq.\eqref{eq. dirac op} for a given, fixed value of the index $n$. From that equation, this involves eight $U$ matrices and nine spinor fields from the neighboring lattice sites, which corresponds to (9*24+8*18)*8 = 360*8 = 2880 input bytes. The $U \times \psi$ matrix-vector multiplications require 1152 floating point operations for complex additions and multiplications. Adding to that the additional preparatory and final additions and multiplications, one obtains for the complete stencil 1464 floating point operations. Hence, the ratio of memory/computation calls is unfavorable, because a lot of data has to be loaded while not so many floating point operations are needed to perform the required matrix times vector multiplication. There is only a small overlap of input variables for a neighboring stencil.

\section{Programmable logic devices advantages}
\label{sec. pld}
In this section we briefly explain what are FPGA devices and highlight these of their features that are important for the main ideas presented in this paper.

FPGA devices are known in the industry and science for years but are most commonly related to specific, low-level, true real-time applications like digital signal 
processing, networking or interfacing between electronic components. This was due to the limited amount of primitive resources and low-abstract development 
methodologies that require manual designing at the level of single flip-flops known as Register-Transfer Level (RTL). 

Nowadays it is no longer the case. Manufacturing process has reached 16 nm, a value similar to the one used for CPUs but first engineering samples of devices 
produced in 7 nm process have been presented, aimed to be publicly available by the end of the year 2018 \cite{Sims:2018}. The number of configurable logic blocks 
contained in a single device has quadrupled since 2012, what is incomparable to the performance increase of the CPUs during the same period of time. 
Not only the amount of resources but also the diversity and the complexity of hardware components that are supporting programmable logic arrays has increased. 
In a single package we can find more than hundred multi-gigabit transceivers (single link with rates up to 58 Gbps), more than 12 thousands Digital Signal 
Processing (DSP) blocks with accumulated processing bandwidth at the level of 21.1 TMACs and memory resources like distributed RAM blocks and up to 8 GB memory 
in a form of stacked silicon in a single 3D integrated circuit package with bandwidth of 460 GBps \cite{Wissolik:2017}.

Development of the design for such a complex device at the RTL level becomes highly challenging. Advancement of the hardware came in pair with the introduction 
of high-level development environments such as Software Defined software family (SDx) and High-Level Synthesis (HLS) from Xilinx for instance \cite{UG1197:2018}. 
Algorithmic parts of the design can now be written as C/C++ and compiled into RTL form using HLS. A set of detailed reports and versatile pragmas give 
the developer control over the translation process from high-level programming language into RTL.

Tremendous amount of resources and fast design methodologies yield a new and powerful component for HPC applications. Most significant change in the paradigm 
of how the application is being developed shifts from optimizing the algorithm to make the best use of a given hardware architecture to designing a hardware 
architecture that will most efficiently compute a given algorithm. Programmable logic presents a number of game-changing features that can make a strong 
impact on how the computing resources for HPC are being designed. The key ones are described below.

Natural parallelism is the most powerful aspect of the FPGAs. One can implement as many instances of some logic component as there are resources available. All those
instances will run in parallel with respect to each other. For instance in Conjugate Gradient algorithm one can implement multiple instances of the module that 
performs matrix times vector multiplication, which is the most time consuming operation. This feature, when used properly, also yields well-scalable solutions.
With the introduction of new devices with higher amount of resources, the overall solution can be easily accelerated by instantiation of additional modules.

Multi-gigabit transceivers are embedded within the FPGA package. Each received data word can be instantly accessed by some processing logic, without interrupts, 
memory transfers or engaging operating system. The latency from the transfer medium to processing module is reduced to minimum. Keeping in mind the number of 
transceivers in most advanced devices and natural parallelization, when data from each link can be processed in parallel, this combination can result in a very
efficient infrastructure for a remote computing node in future HPC architectures. We will develop further this idea in section \ref{sec. multinode}.

Programmable logic has no dependency nor overhead imposed by the operating system, which is required to execute computations on CPUs and GPUs. It is 
therefore also fully resistant to threats or malicious software that could degrade its performance or damage the system.

Logic resources used by a given algorithm components can be dynamically reconfigured to compute another algorithm or even released in order to reduce power usage.
Although this process is rather slow (order of microseconds) and is not suitable for real-time applications, in HPC can be extremely useful for making the best use 
of available resources and power management.

\section{Previous work}
\label{sec. previous work}

There is a continuous development of algorithms and implementations of Lattice QCD kernels for various micro-architectures and HPC solutions. The status is reviewed at the yearly Lattice conference and the most recent review can be found in \cite{Rago:2017pyb}. Most recent progress in the implementations of state-of-the-art solvers on GPU and many-core Xeon Phi accelerators was reported in \cite{joo201716, clark2018pushing, brower2018multigrid} and \cite{Heybrock:2014iga, Heybrock:2015kpy, Georg:2017zua, Kanamori:2017urm} respectively. Obtained performances are of the order of 300 \cite{Durr:2017clx} - 900 \cite{Boyle:2016lbp} GFLOPs single precision for the 64 core Intel's Knights Landing architecture and 500 GFLOPs for double precision and 1.5 TFLOPs or even more for mixed precision on the NVIDIA Volta GPU \cite{clark2018pushing} using multiple right hand sides optimization.

The first FPGA implementation of the conjugate gradient in the context of Lattice QCD was reported in 2005 and 2006 in conference proceedings \cite{DBLP:conf/ipps/CallananNOSG05} and \cite{4100953} and in \cite{phdthesis} in full detail. The Dirac operator was implemented using 2616 double precision floating point operations and the reported performance was up to 1.3 GFLOPs on the ADM-XRC-II development board featuring the Xilinx Virtex-II FPGA, XC2V6000FF1152. This research was not followed up in any way.

\section{Design partitioning}
\label{sec. design_partition}

It is important to carefully select which parts of the algorithm present the highest potential for hardware acceleration. Many considerations should be taken into account, 
therefore it is vital to understand the general mechanism of delegating computations to programmable logic.

\begin{figure}
\centering
\caption{Schematic view at the key components of Zynq MPSoC used to transfer data between PS and PL}
\includegraphics[width=200pt]{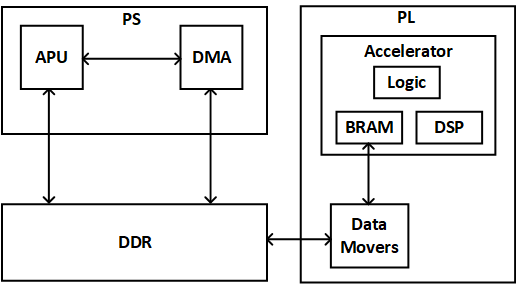}
\label{soc_scm}
\end{figure}

We have chosen a Xilinx Zynq MPSoC (Multi-Processor System-on-Chip) as the device to evaluate the solution (Fig. \ref{soc_scm}). It is composed of two main functional blocks: Processing System (PS) which consists of ARM cores called Application Processing Units (APU) and its hardware entourage as well as Programmable Logic (PL) which are standard FPGA resources. Note that a similar partitioning exists for the Alveo accelerators with the CPU playing the role of the PS and the Alveo FPGA device playing the role of PL. In the following we concentrate on the former setup which we describe in detail. 

The entry point of the design is the software that runs on PS. It executes the code of the algorithm 
and uses external DDR memory as RAM. When the code reaches a call to the hardware accelerator, the required data has to be transported from DDR to the PL.
The software instructs the Direct Memory Access (DMA) controller to fetch particular dataset from the DDR and transport it to the PL through dedicated hardware interfaces.
The data is routed inside the PL by the generated Data Movers to the accelerator that implements a particular software function. The computation results are returned in a reverse
manner to the DDR memory and become accessible to the software on APU.
Therefore, one should look for well isolated fragments that have high amount of computations on limited amount of data with no dependencies to other elements of the algorithm.

Additionally the decision about design partitioning and code structure is driven by the required dataset size and its access pattern from the accelerator: sequential or random. Requesting a data unit from the DDR by the accelerator is slow and inefficient if the entire dataset is large and multiple requests are needed from scattered addresses. For large datasets the DMA should be engaged, which after configuration, streams a particular block of data to the accelerator. Therefore the data in the DDR has to be arranged in contiguous manner and the accelerator has to contain local memory resources: Block RAMs (BRAMs) in which the received data could be stored allowing for random access to all data elements simultaneously. The cost of setting up DMA and transferring data is significant and therefore the partitioning decision should be driven by the goal of accumulating large computational blocks on blocked, non-scattered data. In that case one can profit from streamlined and parallel processing of a single data unit and minimize frequent data transport.

In the case of the CG algorithm there exist a natural candidate to be accelerated in hardware, which is
the vector-matrix multiplication: the multiplication of the spinor field $\psi$ by the Dirac operator $D^{\dagger} D$ in equation \ref{eq. dirac eq}. In such partition the main part of the CG algorithm is
executed on the APU, where the vector scalar products are evaluated and the stopping criterion is evaluated.
Details of an efficient implementation of this idea are described in the next sections.

\section{Software and hardware implementation}
\label{sec. main}

In this section we describe the details of our implementation. We start in section \ref{sec. general} with some general remarks, then we describe the implementation of the data transfer between the DDR RAM and the logic. In section \ref{sec. data mover} we describe the data partitioning within the BRAM blocks and computation parallelization in the logic in section \ref{sec. bram}. In section \ref{sec. stencil} we provide details on the most fundamental part of the kernel function, namely the single stencil computation. We finish this part with remarks on a way of increasing data locality in section \ref{sec. data locality}.

\subsection{Generalities}
\label{sec. general}

In order to easily compare the achieved performance to the standard CPU architectures we impose double precision on all stages of the computation. This comes at a cost of requiring large logic resources to handle complex arithmetics of double precision numbers. A single addition or multiplication of two double numbers takes 14 clock cycles on the FPGA device used in this study. We note that modern GPU implementations of similar algorithms \cite{Clark:2009wm} use various combinations of low precision steps (single or half) in order to increase performance. The impact of this optimization should be investigated in more details in the future but a quick inspection of the generated adder gives some insight. Three DSP hardware blocks are used in order to add two 64 bit floating point numbers. This comes from the fact that the bit-width of the inputs to the block is limited and the adder itself is limited to work on 48 bit values. Reduction of precision, so that the bit-width of the values fits the architecture of the DSP block would result in significant saving of the computing resources.

The great advantage of the FPGA implementation is the fact that we do not need to prepare data structures for contiguous memory access. Once the required data blocks are copied from the DDR memory to the memory blocks of the device data can be accessed in parallel and the matrix-vector multiplications executed simultaneously. Note that for instance the ordering of the real and imaginary parts in memory does not longer play a role, as all the appropriate reshuffling will be executed in hardware data routing, which is computationally costless. Hence no structure of arrays data layout optimization is needed. Note also that usual optimization required on modern CPU processors when the data had to be packed into vectors of the length of registers reaching 512 bits or 8 doubles, is no longer required. Vectors of arbitrary length can be processed in the hardware of a FPGA device with equal ease. In practice we work with abstract types \verb|complex|, \verb|su3_vector|, \verb|su3_matrix| and \verb|su3_spinor|:
\begin{verbatim}
struct complex {
public:
        double r,i;
        [...]
}
struct su3_vector {
public:
        complex v[3];
        [...]
}
struct su3_matrix {
public:
        complex m[9];
        [...]
}
struct su3_spinor {
public:
        su3_vector s[4];
        [...]
}
\end{verbatim}

We experimented with a compressed data structure for the $U$ matrix which is a unitary $3\times3$ complex matrix. The mathematical properties of $U$ ensure that it is enough to store the first two rows, while the third row can be recomputed when needed \cite{DeForcrand:1986inu}. Such solution alleviates the memory bandwidth limitation at the cost of requiring more compute resources. We found however, that the additional computations needed to restore the full matrix prohibit a full pipelining which is more disadvantageous than the gain from the reduced pressure on the data transfer. We defer the discussion of the impact of alternative parametrization of a single $U$ matrix by 8 real numbers to the next publication.

In order to allow the compiler to take advantage of the natural parallelism of FPGA devices it is crucial to store data in PL in such a way that independent data is located in separate BRAM blocks. This limitation corresponds to the fact that in a single PL clock cycle only one memory element can be read from the BRAM block (hardware interface limitation). In the computation of a single stencil one needs eight different $U$ matrices, ideally they should be stored in eight separate BRAM blocks. The set of eight $U$ matrices has a natural partitioning according to the four values of the $\mu$ index. These four sets are fully independent. We further duplicate each of these arrays so that the forward and backward matrices, i.e $n+\hat{\mu}$ and $n$ arguments on Eq.\eqref{eq. dirac op} can be accessed simultaneously. We observe that without this improvement the interval of a single stencil evaluation increases dramatically because the computations await for consecutive reads of data from BRAM memory.  We list the details of the BRAM blocks allocation in section \ref{sec. bram}.

Apart from natural parallelism, FPGA devices offer pipelined processing. The full pipelining in computational block means achieving Initiation Interval (II) of one clock cycle, i.e. the hardware block can accept new input data at each clock cycle. The fixed structure of our computations, namely execution of exactly the same computations for each lattice site, allows for a fully pipelined Dirac matrix multiplication, if only enough logic resources are available on the device.

This is exactly what we observe in section \ref{sec. benchmarks}: the evaluation board which we used for testing does not possess enough memory blocks and hence II for the single stencil calculation block is much longer than one clock cycle. On the other hand, compilation on larger devices equipped with additional Ultra RAM memory (URAM large, embedded memory blocks) shows that full pipelining is possible. Details of the single stencil implementation are described in section \ref{sec. stencil}. 

Hence, an efficient implementation of the CG algorithm on a single FPGA device is limited by the memory type and capacity available on the device. In the next section we describe the benchmarks obtained on the evaluation board where lack of large memory blocks dramatically limit the achievable performance. On contrary, the performance increases three orders of magnitude in larger devices with more memory.

\subsection{Data transmission}
\label{sec. data mover}

The key issue in achieving high performance is the ability of quickly transferring data from the host memory to the accelerator. In the case of the FPGA devices data transfer may or may not be initiated at each call to the accelerated function. 

When the $U$ matrices do not change between consecutive iterations of the solver they can be loaded to the PL only once at the first function call and then kept in PL during the entire execution of the algorithm.
This happens when the entire lattice fits into the memory blocks within the logic or when multiple devices are used and the lattice can be divided into blocks which do fit in the memory blocks.
In the case when the entire lattice does not fit in the memory blocks within the logic it needs to be reloaded in order to enable the computation of all stencils. In either case, fast data transfer is crucial in achieving the shortest time to solution.

Several ingredients are needed to achieve fast data transfer. They include memory allocation on PS side and selection of data transfer mechanism to PL.
In case the computations delegated to hardware are carefully isolated and do not require PS to interact before the final result is calculated, the dataset should be streamed from the DDR to PL at the beginning and back at the end. The fastest streaming can be achieved by assuring that consecutive physical addresses of DDR hold consecutive array elements and no coherency check mechanisms are required. Those two requirements are fulfilled by allocating arrays on the PS side using \verb|sds_alloc_non_cacheable(x)| command.
It has to be accompanied by a proper generation of the Data Movers in the PL, therefore the compiler has to be instructed how the data in the input arguments is organized and accessed. The attribute \verb|PHYSICAL_CONTIGUOUS| of pragma's \verb|SDS data| informs that the data is continuously ordered on the physical level and the attribute \verb| SEQUENTIAL| states that the PL logic will access elements consecutively. Those instructions will cause the compiler to infer DMASIMPLE interface type, which allows the fastest streaming of large memory blocks between DDR and PL. In our case we declare the $U$ matrices in the following way

\begin{verbatim}
#pragma SDS data mem_attribute( U_x_i_in:PHYSICAL_CONTIGUOUS, ...  )
#pragma SDS data access_pattern( U_x_i_in:SEQUENTIAL, ... )
\end{verbatim}
and call the accelerated function providing them as arguments
\begin{verbatim}
void multBatch(double U_x_i_in[vol], ...
\end{verbatim}

One has to take care that all transfer channels between the DDR and PL are used. The evaluation board Xilinx ZU9EG has 8 data channels, whereas the hardware accelerated function has in principle only 6 input arguments: $\psi_{\textrm{in}}$, $\psi_{\textrm{out}}$, $U_x$, $U_y$, $U_z$ and $U_t$. Therefore just for that reason we separated the real and imaginary parts of our input data and sent them to the logic, thus utilizing the full bandwidth.

\subsection{Memory blocks allocation}
\label{sec. bram}

In order to achieve full pipelining the data transferred from the DDR has to be stored in the BRAM blocks in an appropriate way. All calls to memory in the accelerated function have to be inspected. The crucial point here is that one has to make sure that computations which operate on different data and therefore can be executed in parallel have access to the required memory blocks. One has to ensure that all data needed at a given clock cycle reside in different BRAM blocks or registers so that they can be read simultaneously. 

A single BRAM has a fixed data width and capacity, defined by the hardware architecture of the device family. For instance Zynq Ultrascale+ has blocks with 36 kBits capacity and width of 72 bit data words. When storing 64 bit values of the type \verb|double| 8 bits are left unused and inaccessible per entry. Similarly, the remaining capacity cannot be shared with other arrays. Therefore it is important to properly arrange memory resources with the provided mechanisms that allow to instruct the HLS compiler to generate advanced BRAM structures, while letting the developer use standard \verb|array| abstracts.
Pragma \verb|HLS ARRAY_PARTITION| is the simplest way to receive more hardware interfaces to access multiple array elements in parallel by dividing one large array into many smaller with the content distributed to more BRAM blocks, each with its individual interface. Complete partitioning means that all N elements of a given dimension of an array will be distributed into N separate memory instances (BRAMs or registers), each accessible at the same time.

Similar mechanism is employed by \verb|HLS ARRAY_RESHAPE| pragma. Data can be distributed to multiple memory resources but additional vertical control is possible, which allows for a better use of the fixed 72 bits data width. Lets consider a 2-dimensional array of integers (32 bit), where one dimension has size 2. Complete partition of this dimension will generate two separate arrays, housed in two BRAMs. This will result in parallel access however also in a waste of resources because in each array only 32 bits out of 72 bits available will be used by a single element. Complete reshaping of the same array will result in a better arrangement of the elements, by reordering them vertically, so first all available bits of the memory block are used and then additional BRAM is used. In our example, two elements of 32 bits can be packed into one 72 bit data word and therefore only one BRAM is required while still preserving access to both elements at the same time.

\begin{figure}
\centering
\caption{Various schemes of array data arrangement within BRAMs}
\includegraphics[width=200pt]{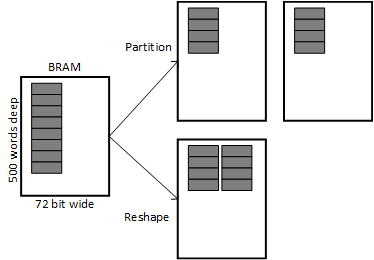}
\end{figure}

In our case we declare the storage of the $U$ matrices on the PL side as (we separate the real and imaginary parts for the reasons of transmission performance, see \ref{sec. data mover})
\begin{verbatim}
double U_x_i[2][vol][9];
double U_x_r[2][vol][9];
\end{verbatim}
and we enforce on these matrices the following properties
\begin{verbatim}
#pragma HLS array_partition variable=U_x_i complete dim=1
#pragma HLS array_reshape variable=U_x_i complete dim=3
#pragma HLS resource variable=U_x_i core=XPM_MEMORY latency=1 uram
\end{verbatim}
which ensures that the two copies of the array are stored separately (complete partitioning in the first argument) and the elements are reordered on the fly according to the third argument, i.e. color indices. The last pragma enforces the variable to be placed in the large URAM memory banks.

We have designed a PL memory infrastructure in a way that all required elements are accessible at a given clock cycle by the computing kernel. Set of reports generated by the HLS tool allows to identify bottlenecks in memory access that stalls computation flow. Using both memory management pragmas on selected arrays and applying data duplication we have achieved II=1 on devices that contain URAM blocks.  

\subsection{Stencil evaluation implementation}
\label{sec. stencil}

\begin{figure}
\centering
\caption{Computation sequence of the stencil solver.}
\includegraphics[width=200pt]{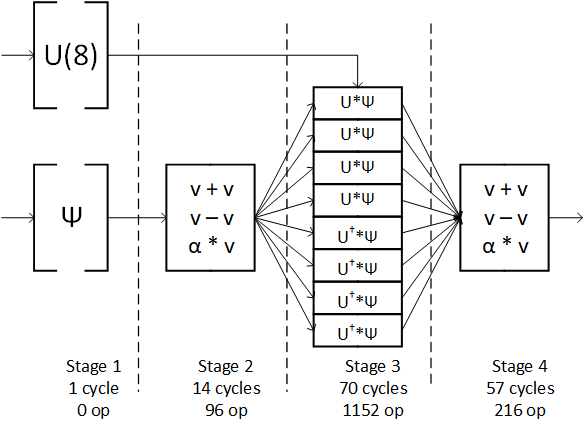}
\label{kernel}
\end{figure}

Single stencil evaluation is implemented as a separate hardware function composed out of 4 main processing stages (Fig. \ref{kernel}). We note that addition and multiplication of two double type variables consumes 14 clock cycles on the VU13P device. This number can differ for different devices and depends on other compile-time constraints.

First (Stage 1) we copy all the necessary data from the BRAM memory blocks to local registers which we declare with the \verb|ARRAY_PARTITION complete| attribute. The process requires only one clock cycle to collect all needed data. It is important to separate all variables so that they can be accessed simultaneously, however we can stick to our abstract types such as the \verb|su3_matrix| type,

\begin{verbatim}
su3_matrix link[8];
#pragma HLS ARRAY_PARTITION complete variable=link dim=1

su3_vector u[8];
#pragma HLS ARRAY_PARTITION complete variable=u dim=1
\end{verbatim}

In Stage 2 we prepare linear combinations of input data. It requires 8 additions and 8 subtractions of \verb|su3_vector| type, that is 96 basic operations on \verb|double|. They are all performed in parallel, therefore the Stage 2 takes 14 clock cycles to provide the result.

Most operations are accumulated in Stage 3, where  \verb|su3_matrix| and \verb|su3_vector| multiplications are performed. Decomposing to basic data type, they require 1152 operations on \verb|double|. The matrix times vector operation is executed in 3 steps: first all nine complex by complex multiplications are evaluated in 28 clock cycles which correspond to a multiplication and an addition of doubles. Then the summation of three complex numbers is performed in 28 cycles corresponding to two double additions. Finally, a rescaling by $\kappa$ requires another 14 cycles, corresponding to a single multiplication of doubles. In total a 5-layer operation cascade is generated and the latency to compute the output \verb|su3_vector| takes $28+28+14=70$ cycles. Note that this is a minimal number of cycles required to analogously perform the multiplication of two \verb|su3_vector|'s, which means that the entire matrix structure is hidden by parallelization. Note also that once the matrix $U$ is in the registers we use it to perform, in parallel, the multiplication of two \verb|su3_vector|, which belong to a single spinor $\psi(n)$. To evaluate the single stencil, 8 such multiplications are required. By using the completely partitioned arrays and instructing the HLS compiler that there are no dependencies between those arrays, we can unroll this loop completely given the required amount of resources and the following instructions are executed in parallel

\begin{verbatim}
for (int i = 0; i < 4; i++) {
#pragma HLS dependence variable=link inter false
[...]                               
 dagger_matrix_times_vector(link[i], u[i],  w[i], uu[i], ww[i]);
}

for (int i = 4; i < 8; i++) {
#pragma HLS dependence variable=link inter false
[...]                               
  matrix_times_vector(link[i], u[i], w[i], uu[i], ww[i]);
}
\end{verbatim}

Finally (Stage 4) we scale and add up all contributions to the final result. Because of the dependencies between consecutive partial results, we had to create a 4-layer operation cascade, which in total takes $57=(4*14)+1$ clock cycles, 4 additions plus one data copy.

Overall, the kernel requires 142 clock cycles and a total of 1464 basic operations to compute the final result since the reception of the input data. It is important to emphasize that, when the kernel is fully pipelined it can accept new input data at each clock cycle and produce the results with latency of 142 cycles, which means there is no dead time in the computation flow, except for startup iteration. Assuming the input data is delivered at requested speed and the clock frequency is 500 MHz the kernel itself is capable to achieve 676 GFLOPs performance. All computations have been decomposed into basic operations and those possible to parallelize have been implemented so. The only factor preventing further optimization is the critical path created by the dependency scheme between partial results.

The implementation is fully scalable. If enough compute resources are available on the device, a second instance of the hardware accelerated kernel can be instantiated. Because the computations for each lattice site are completely independent, the set of lattice sites could thus be divided into two parts and each of them could be associated to one of the kernels, reducing the total computation time by a factor of two.

\subsection{Data locality}
\label{sec. data locality}

In order to increase locality of the implementation we further divided the local lattice into two blocks, by splitting into two halves the data from one dimension. The reason for this optimization is that we gain from the fact that two blocks are completely independent and can be analyzed fully
in parallel if only the device has enough resources. Moreover, higher clock frequencies are possible because we increase locality of the hardware implementation by reducing the number
of calls to distant memory blocks. Greater locality yields much more flexibility to the compiler which can optimize the code better. This optimization requires to store twice a copy of the boundary data for each block separately, which however is negligible compared to the optimization described above when the entire copy of all $U$ matrices is needed to allow parallel access to the data.

\section{Benchmarks results}
\label{sec. benchmarks}

In this section we describe the timings gathered for the CG algorithm. First we comment on the obtained compilation logs focusing on the resource utilization summaries, then we discuss timings and FLOP performances for various hardware configurations. 

\subsection{Compilation details}
\label{sec. resources}

\begin{table}
\begin{center}
\caption{Resource consumption versus parallel execution. Abbreviations in the column names have the following meaning: BRAM - Block RAM, DSP - Digital Signal Processing slice, FF - Flip-Flop unit, LUT - Look-up Table unit, URAM - Ultra RAM. The percentages assigned for the ZU9EG device and Alveo refer to the fraction of the total resources used for the project. The last column indicates whether the resources were consumed by the kernel only or by the full solution. \label{tab. resources}}
\begin{tabular}{|c|ccccccc|c|}
\hline
Device & Latency&Interval&BRAM&	DSP&	FF $[10^6]$ &	LUT $[10^6]$&	URAM & compiled \\
 & & & & & & & & unit \\
\hline
XCVU13P & 142&	1&	508&	6960&	1.58&   0.99&	696 & kernel\\
XCVU13P & 151&	2&	448&	4320&	0.97&	0.64&	696 & kernel\\
XCVU13P & 151&	2&	428&	3480&	0.83&	0.57&	696 & kernel\\
XCVU13P & 162&	4&	412&	1740&	0.47&	0.35&	696 & kernel\\
\hline
ALVEO U250 & 138& 1& 612 (30\%)& 8328 (72\%)& 1.13 (41\%)& 0.74 (55\%)& 696 (54\%)& kernel\\
ALVEO U250 & 138& 1& 774 (39\%)& 8332 (72\%)& 1.28 (46\%)& 0.84 (62\%)& 696 (54\%)& full\\
\hline
ZU9EG & 250&	120&	1388 (76\%) & 546 (21\%) &	0.13 (24\%) & 0.09 (34\%) & $-$ & kernel\\
ZU9EG & 250&	120&	1735 (95\%) & 546 (21\%) &	0.17 (31\%) &	0.11 (40\%) & $-$ & full\\
\hline
\end{tabular}
\end{center}
\end{table}

In table \ref{tab. resources} we report on the amount of resources used for the compilation of the entire project on various devices. We specify the total number of particular architectural blocks: Block RAM,  Digital Signal Processing slices, Flip-Flop units, Look-Up Table units and Ultra RAM needed to compile either the accelerated kernel only or the entire solution containing also the data moving infrastructure.

The first four rows are generated for Xilinx VU13P, at the moment the FPGA with highest amount of resources. We have successfully compiled the design and analyzed the reports in order to provide prospects of how our kernel would perform on high-end devices. Consecutive two entries are results from Alveo U250 accelerator board for which we have compiled the design and emulated it with Quick EMUlator (QEMU - part of Xilinx programming suite). Last entries show the results of the design compiled and tested on hardware using Xilinx ZU9EG. For the reference, in table \ref{tab. full resources} we present available resources on each of these devices.

\begin{table}
\begin{center}
\caption{Resource availability on different devices discussed in the text. Column names are the same as in table \ref{tab. resources}. Data from Xilinx datasheets: \cite{resources},\cite{resources2}. \label{tab. full resources}}
\begin{tabular}{|c|ccccc|}
\hline 
Device & BRAM&	DSP&	FF $[10^6]$ &	LUT $[10^6]$&	URAM\\
\hline
ZU9EG&	1824&	2520&	0.55 &   0.27 &	-\\
ALVEO U250 &2000&	11508&	2.75&	1.34&	1280\\
VU13P &5376&	12288&	3.46&	1.73&	1280 \\
\hline
\end{tabular}
\end{center}
\end{table}

The compilation process can be controlled by a set of parameters as \verb|#define| in the include files and arguments in \verb|#pragma| statements. One can control
\begin{itemize}
\item the problem size by setting the dimensions of the lattice, which also fixes the sizes of generated arrays ($\psi(n)$, $U_{\mu}(n)$ as well as nearest-neighbor tables), 
\item the partitioning/reshape factors of the data arrays, 
\item number of instances of computing kernels by \verb|HLS allocation instances|
\item parallelization factors with \verb|HLS unroll factor|
\end{itemize} 
Using those parameters one can find a balance between resource usage and achieved Iteration Interval.

For the following results we used HLS version 2018.2. There is a noticeable change between this and the previous version of the compilation suite which enforces the tool to produce results with lower II rather than minimize resource usage.

From the resource usage collected in table \ref{tab. resources} one can derive many conclusions. The most important one is that it is possible to generate a kernel capable to compute single stencil within 142 cycles and maintain II=1 allowing fully pipelined designs. Example of emulation report of such kernel is shown in figure \ref{fig. report}. Middle-range devices do not have enough internal memory resources in order to properly distribute array elements for the kernel to compute new results at each clock cycle. Therefore, the memory usage is crucial while achieving II=1. Both Alveo U250 and VU13P have enough resources to accommodate fully pipelined kernel instance, although a difference in consumption can be noted. This comes from the fact that the design for Alveo was compiled with clock frequency 300 MHz and for VU13P it was 500 MHz. In such case, the compiler optimization process has selected to use more FF and LUT for arithmetic operations at the expense of DSP. On ZU9EG 76\% of memory is used (the rest has to be reserved for Data Movers) but only 21\% of DSP blocks, which means that most of the computing resources are left unused due to lack of parallel memory interfaces and duplicated data units.

The application also requires large datasets to be held in internal memories. Devices that contain URAMs have a great advantage over devices with a number of BRAMS. This is because access to a single large memory allows achieving better timing and avoids long signal propagation paths in generated routing.

One can also notice lack of improvement when the unroll factor is not consistent with the loop trip count. Additional resources reserved by subsequent instances of the kernel are not fully used because the flow control logic has to wait until uneven kernels finish.

\subsection{Timing analysis}
\label{sec. timing analysis}

\begin{figure}
\centering
\caption{Emulation timeline of the kernel. The left column shows the case of II=16 and the right column shows the kernel achieving initiation interval of 1 clock cycle, both obtained on the ALVEO U250 device. Blue signals correspond to reception of the input data sets, red to the transmission of the output results and lower green signal shows an execution of one the kernel components. The set of input data was reduced to 32 sites in order to improve readability of the figure. Figure is a screenshot of the QEMU capture. \label{fig. report}}
\includegraphics[width=400pt]{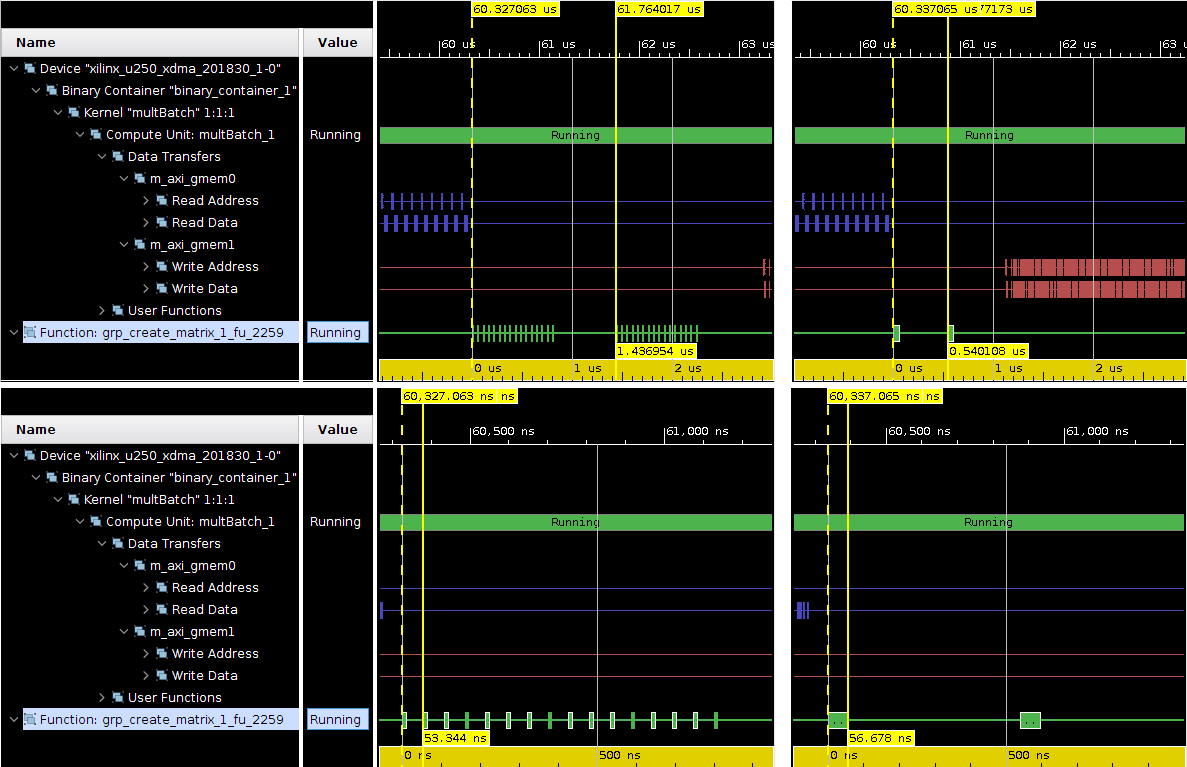}
\end{figure}

Figure \ref{fig. report} shows a sample of results of the emulation of our kernel on the Alveo U250 device. In order to increase the readability of the figure we have decreased the problem size to $V=2^3 \times 4 = 32$ lattice sites. The left part of the figure shows which channels were monitored during the emulation. On the horizontal axis of the right part of the figure emulated time is shown in micro or nano seconds. The most interesting signal marked in green in the lower part of the screenshots is the function \emph{grp\_create\_matrix} which is a component of our kernel. The corresponding timeline on the rights shows at which clock cycle that particular function is executed. The remaining signals, blue and red, show the data transfer to and from the kernel. The right part of the figure has two columns: the left column shows results for an emulation where the initiation interval for the kernel was set to 16 clock cycles, whereas the right column corresponds to an initiation interval of 1 clock cycle. Because, as we described above in section \ref{sec. data locality}, we divide the analyzed lattice into two parts, the signal has two parts, each consisting of 16 calls to the kernel. The upper row shows the timeline on a larger time scale providing an overview of the entire execution of the program. When II=1 the calls form a continuous band as at every clock cycle an execution of that function is initiated. The lower row shows consecutive calls to the kernel in more details. One can see that when II=16 the delay between the execution of the \emph{grp\_create\_matrix} is 53.344 ns which corresponds to exactly 16 clock cycles at 300 MHz. Correspondingly, in the case of II=1, 16 calls to that function take 56.678 ns which is equal to 16 clock cycles + 1 additional clock cycle. Although for the timeline we had to chose a particular function, the same conclusions must remain true for the entire kernel, because each kernel call contains exactly one call to the \emph{grp\_create\_matrix} function.

\subsection{Problem size limits}
\label{sec. size limits}

In this section we briefly comment on the maximal problem size that fits in the currently available FPGA devices using our implementation. The problem size is given by the dimensions of the four-dimensional lattice where the gauge and spinor fields are defined. We parametrize them with two integers $L$ and $T$, $L$ being the number of lattice sites in three spatial directions and $T$ being the extend in the time direction. The total number of lattice sites is given by $V=L^3 \times T$ and is equal to the number of the calls to the hardware accelerated single stencil computation kernel.

In table \ref{tab. resources size} we report on the dependence of FPGA resources used as a function of the problem size. The first entry corresponds to the maximal problem size which was run on the Xilinx ZU9EG device.  The limited memory resources available on that device allow to use a lattice of size $V=6^3 \times 8$. We remind that the lattice is divided into two sublattices, $6^3 \times 4$ each in order to increase data locality. Subsequent entries correspond to problem sizes which were compiled for the Xilinx VU13P device and fit into that device memory. We notice that the consumption of compute resources does not change as the problem size increases, only the amount of used memory blocks increases.

\begin{table}
\begin{center}
\caption{Resource consumption versus problem size. Column names are the same as in table \ref{tab. resources}. All lattice sizes fit in the largest available VU13P device. \label{tab. resources size}}
\begin{tabular}{|ccccccc|}
\hline 
L&T&BRAM&	DSP&	FF $[10^6]$ &	LUT $[10^6]$&	URAM\\
\hline
6&	8&	508&	6960&	1.58&   0.99&	696\\
8&	8&	508&	6960&	1.58&	0.99&	888\\
8&	10&	508&	6960&	1.58&	0.99&	888\\
8&	12&	508&	6960&	1.58&	0.99&	1080\\
\hline
\end{tabular}
\end{center}
\end{table}

\subsection{Performance benchmarks}
\label{sec. timings}

\subsubsection{Methodology}

The timings are collected from on-hardware execution, full system emulations and compilation logs. We count the number of cycles actually needed for the execution of particular steps of the calculation (reading data in from DDR, evaluating all stencils, writing data out to DDR). Assuming that the input data does not need to be loaded from the DDR memory for each consecutive call of the accelerated function, the duration of the computations can be calculated knowing the degree of parallelization and the following parameters: the interval $\delta$ and latency $\tau$ of the compute kernel, the clock frequency $\nu$, the number of lattice sites $V$ to be processed and the number of FLOPs per lattice site $f$,

\begin{equation}
\textrm{performance} = V \times f \times \nu / ( V \times \delta + \tau )
\label{eq. formula}
\end{equation}

In order to estimate the number of FLOPS per lattice site $f$, we added appropriate counters for all arithmetic operations of the basic types (\verb|int|, \verb|double| and \verb|complex|) used in our implementation and explicitly counted all floating point operations, obtaining 1464 FLOPs per lattice site.

\subsubsection{Results}

In table \ref{tab. flops} we report on the obtained performances in GFLOPs. The numbers come from the inspection of the trace data captured during algorithm emulation or execution in hardware and application of Eq.\eqref{eq. formula}. Results for the VU13P device have been extrapolated from compilation logs.

\begin{table}
\begin{center}
\caption{Achieved performance in GFLOPs. \label{tab. flops}}
\begin{tabular}{|c|c|c|c|}
\hline
Device & Initiation & Clock frequency & Performance \\
& Interval & [MHz] & [GFLOPs] \\
\hline					
VU13P & 1&500&676\\
\hline
ALVEO U250 & 1 & 300 & 405 \\
\hline
ZU9EG& 120&150&1.82\\
\hline
\end{tabular}
\end{center}
\end{table}

Maximum performance can be achieved with the design that has the lowest II. That is possible on the VU13P and Alveo U250 devices which have enough memory and DSP blocks available. Assuming no data transfer is required and the clock frequency is at the level of 500 MHz (kernel timing closure obtained) the generated computing logic reaches 676 GFLOPs. Correspondingly for the emulated kernel on the Alveo device at 300 MHz we obtained the performance of 405 GFLOPs. The performance scales linearly with the amount of fully utilized kernels. 

Due to limited memory resources we have achieved only 1.8 GFLOPs on the available ZU9EG. Not only high number of clock cycles for II but also lower clock frequency (Virtex Ultrascale+ is the high-end devices family and contains URAMs) are the reason for slower design.

As it was explained in Section IV, the kernels have to wait until all data is transported from the DDR to PL. Including the time required by this process, we achieved 1.3 GFLOPs on the available platform. The ratio between with and without data transport is strongly dependent on the hardware infrastructure available on the FPGA chip itself and the hardware platform it is mounted on. Our ZU9EG device has 4 hardware interfaces between DDR and PL. Using Virtex Ultrascale+ devices, one can profit from tens of embedded multigigabit transceivers to stream data into the device. Although it will require to design a proprietary data distribution system, we expect to achieve much better ratio than in the case of the proof-of-concept platform.

\section{Prospects for multi-node systems}
\label{sec. multinode}

It is clear from the results described in previous sections that practically interesting problem sizes can
be dealt with only using a multi-node system. It also follows that FPGA devices cannot be used as simple accelerators because the data transfer between DDR memory and the logic severely limits the achieved performance. Therefore the only viable solution is to connect the hardware logic blocks directly together bypassing the control of the host processing unit. This can be achieved using the built in transceivers and a simple send/receive infrastructure implemented directly in the logic. 

The initial problem would be then divided into blocks each of which would be processed by a single FPGA device. The data would be loaded once at the beginning and kept in the logic between consecutive iterations of the algorithm. The interconnect infrastructure would be used to exchange boundaries between adjacent blocks. In such a setup the calculation of the norm of the residue vector could also be executed in the logic and only the norm would be returned to the control host processing unit, where the stopping criterion would be evaluated. 

Our preliminary results point that an HPC solution based entirely on FPGA devices could outperform current installations at a considerably smaller electric power consumption.

\section{Conclusion}
\label{sec. conclusions}

In this paper we have presented our implementation of the Conjugate Gradient algorithm as used in the Monte Carlo simulations of Quantum Chromodynamics. We described various optimizations for the specific micro-architecture of the FPGA devices. We benchmarked our solution and found out that the obtained performance is comparable with the one obtained on modern CPU units. 
Our computing kernel is fully pipelined and can achieve 676 GFLOPs with double precision.
We conclude therefore that FPGA devices can be (taking into account their development pace - should be) considered as a viable solution for HPC systems as far as the specific application considered in this paper is concerned. Scalability of this solution still has to be demonstrated and benchmarked, we have however pointed out a direction which seems to be particularly encouraging. 

\appendix

\section{Conventions}
Dirac matrices are given by
\begin{equation}
\gamma_0 = \left( \begin{array}{cccc} 
0 & 0 & 0 & i \\
0 & 0 & i & 0 \\
0 & -i & 0 & 0 \\
-i & 0 & 0 & 0 
\end{array} \right), 
\gamma_1 = \left( \begin{array}{cccc} 
0 & 0 & 0 & -1 \\
0 & 0 & 1 & 0 \\
0 & 1 & 0 & 0 \\
-1 & 0 & 0 & 0 
\end{array} \right),
\gamma_2 = \left( \begin{array}{cccc} 
0 & 0 & i & 0 \\
0 & 0 & 0 & -i \\
-i & 0 & 0 & 0 \\
0 & i & 0 & 0 
\end{array} \right), 
\gamma_3 = \left( \begin{array}{cccc} 
0 & 0 & 1 & 0 \\
0 & 0 & 0 & 1 \\
1 & 0 & 0 & 0 \\
0 & 1 & 0 & 0 
\end{array} \right).
\end{equation}
Additionally, we have
\begin{equation}
\gamma_5 = \left( \begin{array}{cccc} 
1 & 0 & 0 & 0 \\
0 & 1 & 0 & 0 \\
0 & 0 & -1 & 0 \\
0 & 0 & 0 & -1 
\end{array} \right).
\end{equation}

\section*{Acknowledgment}
The Authors thank K.K. for many valuable discussions. 

This work was in part supported by Deutsche Forschungsgemeinschaft under Grant No. SFB/TRR 55
and by the polish NCN grant No. UMO-2016/21/B/ ST2/01492, by the Foundation for Polish Science grant no. TEAM/2017-4/39 and by the Polish Ministry for Science and Higher Education grant no. 7150/E-338/M/2018.

The project could be realized thanks to the support from Xilinx University Program and their donations.

\bibliographystyle{elsarticle-num}
\bibliography{references}

\end{document}